 \documentclass[preprint]{elsarticle}

\usepackage{graphicx}
\usepackage{amsmath}
\usepackage{amsfonts}
\usepackage{epsfig}
\usepackage{hhline}
\usepackage{soul}
\usepackage{tabularx,pifont,multirow}
\usepackage{enumitem}
\usepackage{colordvi}

\newcommand\trule{\rule{0pt}{4.5ex}}

\newcommand{\VA}[3]{\ifthenelse{\equal{#2}{#3}}
{\ensuremath{#1\pm#2}}{\ensuremath{#1\,^{+#2}_{-#3}}}}

\def\slash#1{\setbox0 = \hbox{$#1$}#1\hskip - \wd0\hbox to\wd0{\hss\sl/\/\hss}}

\newcommand{\Kz}{\ensuremath{K^0}}
\newcommand{\Kzb}{\ensuremath{\bar{K}^0}}

\newcommand{\GS}{\ensuremath{\gamma}}

\begin{document}

%

\title{Consistent Probabilistic Description of the Neutral  Kaon System}

\author[jb]{J.~Bernab\'eu}

\author[nm]{N.E.~Mavromatos}

\author[jb]{P.~Villanueva-P\'erez}

\address[jb]{Department of Theoretical Physics, Universitat de Val\`encia, E-46100 Burjassot (Valencia) and 
IFIC, Universitat de Val\`encia-CSIC, E-46071, Paterna (Val\`encia), Spain}

\address[nm]{Theoretical Particle Physics and Cosmology Group, Department of Physics, \\ King's College London, Strand, London WC2R 2LS, UK }

\bigskip 

\begin{abstract}

The neutral Kaon system has both CP violation in the mass matrix and a 
non-vanishing lifetime difference in the width matrix. This leads to
an effective Hamiltonian which is not a normal operator, with
incompatible (non-commuting) masses and widths. In the Weisskopf-Wigner
Approach (WWA), by diagonalizing the entire Hamiltonian, the
unphysical non-orthogonal "stationary'' states $K_{L,S}$ are obtained. 
These states have complex eigenvalues whose real (imaginary) part 
does not coincide with the eigenvalues of the mass (width) matrix.
In this work we describe the system as an open Lindblad-type quantum
mechanical system due to Kaon decays. This approach, in terms of density matrices for initial
and final states, provides a consistent probabilistic description,
avoiding the standard problems because the width matrix becomes a composite operator not included
in the Hamiltonian.
We consider the dominant-decay channel to two pions, so that one of the Kaon states
with definite lifetime becomes stable.
This new approach provides results for the time
dependent decay rates in agreement with those of the WWA.

\end{abstract}


\maketitle

Neutral Kaons is a fascinating physical system that, due to its peculiar and at the time paradoxical behaviour in many respects, has 
lead to important discoveries, thereby triggering 
 an enormous interest for its study. 
 It is the first physical system where CP violation has been observed in the two-pion $K^0 \rightarrow 2\, \pi$ decay channel~\cite{cpviol},  
with the relevant experimental studies continuing up to date~\cite{cplear,na48,ktev} and extended to entangled neutral Kaon states in meson $\phi$ factories~\cite{dafne}. Moreover, neutral Kaons  have also been used as a probe of fundamental symmetries, such as CPT invariance~\cite{cpt}, 
and deviations from the standard quantum mechanical behaviour. The latter may be induced by quantum gravity fluctuations  appearing  as a ``decoherening'' environment, leading to an open system (Lindblad-type~\cite{lindblad}) formulation~\cite{marinov,ehns,decoh,dolgov,lopez,huet,omega}.

The standard description of the neutral-Kaon system follows the Weisskopf-Wigner Approach (WWA)~\cite{WWA,Lee} for unstable particles using the non-Hermitian Hamiltonian
\begin{equation}\label{width}
\widehat{\mathcal{H}} = \widehat{\mathcal{M}} - i \, \frac{\widehat{\Gamma}}{2} ~.
\end{equation}
However, the simultaneous presence of CP violation in the mass matrix $\widehat{\mathcal{M}}$ 
and a difference of lifetimes in the antihermitian matrix $i\widehat{\Gamma}/2$ leads 
to a quantum incompatibility between $\widehat{\mathcal{M}}$ and $\widehat{\Gamma}$,
$\big[ \widehat{\mathcal{M}}, \widehat{\Gamma} \big] \ne 0$ (see discussion below), \emph{i.e.} one cannot define states of definite mass and lifetime simultaneously, because $\widehat{\mathcal{H}}$ is not a normal operator.
In this situation, the eigenstates $K_L$ and $K_S$, 
obtained by a non-unitary 
diagonalization of $\widehat{\mathcal{H}}$, 
lack physical meaning and their non-orthogonality prevents a consistent probabilistic description of
this and 
any other system with a Hamiltonian  which is not a normal operator\footnote{We mention here that the issue of the physical 
meaning of $K_{L,S}$ has also been addressed within an S-matrix formalism, instead of a time-dependent approach~\cite{stodolsky,pilaftsis}.
Reference~\cite{pilaftsis} corrects the earlier treatments of Ref.~\cite{stodolsky}, where $K_{L,S}$ had been 
considered as physical poles of the propagator, by demonstrating the non-factorizability of the propagator matrix for this coupled system.}. 
This is the central point of our discussion for neutral Kaons, in contrast to the $B_d$ meson case, in which  the non-Hermitian $\widehat{\mathcal{H}}$ is a normal operator in a very good approximation.


The lack of a proper probabilistic interpretation of the neutral Kaon system has been 
addressed previously~\cite{alvarez,silva} by distinguishing 
the ket and bra Hilbert spaces, so that the non-orthogonality of the $K_L$, $K_S$ states is avoided with the use of these states and their duals. 
In our treatment, we do not make use of those unphysical states by considering not only the 
dynamics of pure initial states but also the inclusion of the final decay products, which will lead to a time evolution to mixed states.

In \cite{open1, hiesmayr} a suggestion has been made to view a decaying quantum system 
as an open system interacting with an appropriate ``\emph{environment}'' obtained by enlarging the original Hilbert space by states representing 
the decay products. The time
evolution of such a  system can be described by an effective Hermitian Hamiltonian, essentially $\widehat{\mathcal{M}}$ above, and an additional  dissipative 
term of Lindblad form (\emph{dissipator})~\cite{lindblad}. 
As shown in \cite{hiesmayr}, 
the non-Hermitian part of the Hamiltonian in the WWA, associated with the  
particle decay width operator $\widehat{\Gamma}$, can be incorporated
into the dissipator of the enlarged space via a specific Lindblad operator $\mathcal{B}$. 
That work, however, was only 
applied to the trivial case of the decay of a single particle and it did not address 
this method to the interesting case of the neutral Kaon system with
its problem of the incompatibility of $\widehat{\mathcal{M}}$ and $\widehat{\Gamma}$.
This important point is also in contrast with Ref.~\cite{open1} which, although
dealing with the neutral meson systems, they study $K$ and $B_d$ mesons on equal footing,
based simply on the non-Hermiticity of the Hamiltonian. Moreover, they use
explicitely the unphysical non-orthogonal $K_{L,S}$ basis, which prevents a consistent
probabilistic interpretation.
This will be the focus of attention of the present article.

To understand in simple terms the logic behind this open-quantum-system formalism for decaying systems, 
 we first concentrate our attention on the evolution equation  
for the initial density matrix,  
$\rho =  |\Psi \rangle \langle \Psi |$, restricted to the $|K^0\rangle|$ $|\bar{K}^0\rangle|$ system: 
\begin{equation}\label{commutator}
{\dot \rho} = - i \widehat{\mathcal{H}} \, \rho + i \rho \widehat{\mathcal{H}}^\dagger = -i \Big[  \widehat{\mathcal{M}}, \rho \Big] -  \frac{1}{2} \Big\{\rho, \widehat{\Gamma} \Big\}~,
\end{equation}
where ${\dot \rho}$ denotes time derivative. This equation can be \emph{formally} obtained from the Schr\"odinger equation for the state vector $|\Psi \rangle$ with the \emph{non-Hermitian} Hamiltonian $\widehat{\mathcal{H}}$. In Eq.~(\ref{commutator}), $\widehat{\Gamma}$ is viewed as a single quantum mechanical operator. The anti-Hermitian part of $\widehat{\mathcal{H}}$ leads to the anticommutator term in the right-hand-side of this evolution equation. 
As a consequence, the description of the system in terms of pure states, for which ${\rm Tr}\rho^2 = {\rm Tr}\rho$, breaks down at time $t > 0$. This can be readily shown by calculating the rate of the Von-Neumann entropy $\mathcal{S} = - {\rm Tr}\big(\rho \, {\rm ln}\rho\big) $ using the evolution (\ref{commutator}), where only the anticommutator part contributes. 

As a consequence of the restriction to the initial Hilbert space ignoring the decay products, one has ${\rm Tr}{\dot \rho}(t)  \ne 0$.
To restore Unitarity~\cite{bell} we include the final states when taking the trace ${\rm Tr} \rho $ through
a mapping from the initial Hilbert space to the final one (decay products): $H_i \to H_f$. This mapping is implemented~\cite{hiesmayr} by the transition operator ${\mathcal B}$, which is related to $\widehat{\Gamma}$ via:
\begin{equation}\label{defwidth1}
{\mathcal B}^\dagger {\mathcal B} = \widehat{\Gamma}~.
\end{equation}If $\{ f_k \}$ denotes an orthonormal basis in $H_f$, and $\{ \varphi_j \}$ denotes the corresponding orthonormal basis in $H_i$ (orthogonal to $\{ f_k \}$ ), then one may write:
\begin{equation}\label{defB}
\mathcal{B} = \sum_{k=1}^{d_f}\, \sum_{j=1}^{d_i} \, b_{kj} \, |f_k \rangle \langle \varphi_j | ~,
\end{equation}
where $d_f = {\rm dim} H_f$, and $d_i = {\rm dim} H_i$. 
The width operator $\widehat{\Gamma}$ is thus a positive definite self-adjoint operator with non-negative eigenvalues. The latter can include possible zero eigenvalues, corresponding to stable states. This is to be contrasted with the corresponding expression given in \cite{hiesmayr} and will have important consequences for the neutral-Kaon system. 

The operator ${\mathcal B}$ can be considered as a sort of ``\emph{environment}'' operator 
from the point of view of the initial state Hilbert space, and the evolution (\ref{commutator}) can be replaced now by  
an appropriate Lindblad evolution~\cite{lindblad}, with $\rho$ spanning the combined initial ($H_i$) and final ($H_f$) Hilbert spaces, $H_{tot}\equiv H_{i}\bigoplus H_f$. The Lindblad evolution can be understood as follows: 
in view of (\ref{defwidth1}), the simple commutator structure 
of (\ref{commutator}) in the conventional WWA~\cite{lopez} 
will now be replaced by an appropriate \emph{quantum } ordering  
of the constituent operators $\mathcal{B}, \mathcal{B}^\dagger$ 
and $\rho$ in such a way that the time evolution has the following properties~\cite{lindblad}: (i) preserves the complete positivity of the density matrix operators 
at any time, \emph{i.e.}, the fact that their eigenvalues are positive or zero, 
so that the concept of probabilities associated with 
the eigenvalues of these operators makes sense, 
(ii) ensures the conservation of the total probability through ${\rm Tr} \rho = 1$, 
including the \emph{final} states (decay products) and (iii) implies 
increase of the entropy (of quantum-mixed states).

Our density matrix $\rho$ in the total Hilbert space $H_{tot}\equiv H_{i}\bigoplus H_f$ is:
$\rho=
\left(
\begin{array}{cc}
 \rho_{ii'} & \rho_{if} \\
 \rho_{fi} & \rho_{ff'} 
\end{array}
\right)$, where Hermiticity of $\rho$ is fulfilled in blocks and 
the subindices $ii'$ ($ff'$) run over the initial (final) states. 
We have for the dimension of the relevant Hilbert spaces:
\begin{equation}\label{cond} {\rm dim} \, H_i \, < \infty ~, \quad {\rm  and} \quad  
{\rm dim} \, H_f \, \ge r =  {\rm dim }H_i - n_0~,
\end{equation}
with $n_0$ the degeneracy of the eigenvalue zero of the width operator. The evolution equations for the density matrix $\rho$ in the $H_{tot}$ Hilbert space are then described by the Lindblad form~\cite{lindblad}:
\begin{equation}
{\dot \rho} =-i\big[{\cal H},\rho\big]-\dfrac{1}{2}\big({B}^\dagger{B}\rho+ \rho{B}^\dagger{ B}-2{B}\rho{B}^\dagger\big)~,
\label{lind}\end{equation}
with 
\begin{equation}\label{hamilt}
{\cal H}= {\cal H}^\dagger = 
\left(
\begin{array}{cc}
 \widehat{\mathcal{M}} & 0  \\
 0 & 0 
\end{array}
\right), \quad B = \begin{pmatrix} 0 \quad  0 \\ \mathcal{B} \quad 0 \end{pmatrix},
\end{equation}
in total Hilbert space. The new formulation of the time evolution 
on the enlarged space has a hermitian
Hamiltonian and is probability conserving. The complete positivity, that is guaranteed by construction in the Lindblad formalism~\cite{lindblad}, ensures that this
feature characterizes the decaying quantum system, exactly  as it happens in systems with Hermitian Hamiltonians.
This is an effective quantum mechanical approach where the decay is accounted for by the non-Hamiltonian Lindblad environmental operators in Eq.~(\ref{lind}). 
The environment in our approach is not an external agent, unlike the situation encountered in Quantum Field Theory systems at finite temperature~\cite{Millington:2012pf}. 
It is an open question whether the effects of the decay can be reproduced by field theoretical source terms.

We would like to 
discuss here the application of this Lindblad open-system formulation of particle decay 
to physically realistic systems, such as neutral Kaons, which are known to exhibit 
CP violation and non-zero width difference $\Delta \Gamma \ne 0$. Contrary to WWA and the dynamics given by Eq.~(\ref{commutator}), the open-system formalism is applicable in terms of the transition operator $B$ (\ref{hamilt}), irrespective of the commutativity of the composite $\widehat{\Gamma}$ operator  with $\widehat{\mathcal{M}}$. In this respect, the Lindblad dynamics (\ref{lind})  
for the decay is 
particularly interesting for neutral Kaons. Other neutral mesons, such as $B_d$-$\overline{B}_d$ systems, are characterized by 
very small width differences between the physical eigenstates, practically $\Delta \Gamma \simeq 0$, for which the non-Hermitian Hamiltonian is a normal operator and the 
WWA solution by diagonalization of the entire hamiltonian $\widehat{\mathcal{H}}$ is satisfactory.

For this discussion, we focus our attention from now on to two-state unstable systems. 
We can write  Eq.~(\ref{lind}) 
as :
{\small \begin{eqnarray}\label{eq:lind}
 \begin{pmatrix}
 \dot{\rho}_{ii'} & \dot{\rho}_{if} \\
 \dot{\rho}_{fi} & \dot{\rho}_{ff'} 
\end{pmatrix} = 
\begin{pmatrix}
  -i\big[\widehat{\mathcal{M}},\rho_{ii'}\big]-\dfrac{1}{2}\big\{
\widehat{\Gamma},\rho_{ii'}\big\} & \qquad -i\widehat{\mathcal{M}}\rho_{if}-\dfrac{1}{2} \widehat{\Gamma}\rho_{if} \\
 \trule i\rho_{fi}\widehat{\mathcal{M}}-\dfrac{1}{2}\rho_{fi} \widehat{\Gamma} 
 & \qquad \mathcal{B} \rho_{ii'} \mathcal{B}^{\dagger}
\end{pmatrix}. \nonumber \\
\end{eqnarray}}
One notes the following: (i) 
the new dynamical behavior of the final state coupled to the initial one, 
with effects which cannot be described in general by $\widehat{\Gamma}$ only and the explicit form of the Lindblad operator $\mathcal{B}$ is needed; 
(ii) the formally identical structure  of 
the equation for the time evolution of the initial-state density matrix, which is uncoupled to the final states,  to that of Eq. (\ref{commutator}), 
as a result of the anticommutator  $\{ \widehat{\Gamma}, \rho_{ii'}\}$; however, 
in this description, the last term is originated by an ``external'' agent and it is not included
in the hamiltonian of the two body system itself, therefore this term is responsible of the evolution from pure to mixed states
in the sense of ${\rm Tr}\rho (t)^2 \ne {\rm Tr} \rho(t)$ at a time $t > 0$;
(iii) the uncoupled dynamical behavior of $\rho_{if}(t)$, so that it is consistent to 
take $\rho_{if}(t) =0$, if there is no initial ($t=0$) mixed component between initial and final states; this implies that the description of the time evolution and decay is expressed in terms of initial and final density matrices only.
Due to the separate treatment of commutator and anticommutator terms in the initial submatrix of Eq.~(\ref{eq:lind}), the non commutativity
of $\widehat{M}$ and $\widehat{\Gamma}$ is not an issue, avoiding $|K_L\rangle$ and $|K_S\rangle$ states to be used explicitly in the time
evolution of the system.

The reader should recall once more that  in the Lindblad approach, total probability conservation for the density matrix, including the decay products, 
is guaranteed by construction, \emph{i.e.}, ${\rm Tr}\rho_{ii} + {\rm Tr} \rho_{ff} = 1$ for any $t$, so that Unitarity~\cite{bell} is implied by the simple relation:
$\dfrac{d Tr(\rho_{ii}(t))}{d t}=-\dfrac{d Tr(\rho_{ff}(t))}{d t}$. This relation can be verified explicitly for the solutions we obtain here for the case of the neutral Kaon system.

For the neutral Kaon $\Kz-\Kzb$ system we incorporate properly CP violation and the dynamics of its dominant decay to two pion final states. We  use the $|K_{1,2}\rangle $ basis for the Kaon states 
defined as:
{\small \begin{eqnarray} \label{k1k2}
|K_1\rangle  =  \dfrac{1}{\sqrt{2}} \big(|\Kz\rangle - |\Kzb\rangle \big)~, \quad 
|K_2\rangle =  \dfrac{1}{\sqrt{2}} \big(|\Kz\rangle + |\Kzb\rangle \big),
\end{eqnarray}}which, as we show below, is a convenient choice in which the width operator is diagonal. 

 The existence of a \emph{dominant} decay channel in the neutral Kaon system, $\pi\pi$ in isospin $I=0$ 
as dictated by the $\Delta I=1/2$ rule,
implies via  Eq.~(\ref{cond}), that  $n_0 =1$, which is correct, given that there 
is only one zero eigenvalue in the spectrum of $\widehat{\Gamma}$. 
%
Ignoring CPT Violation and CP violation in the decay, the choice of a real $\mathcal{B}$ leads to 
 the result that the $K_{1,2}$ states are the ones with definite lifetimes, so that 
the  width operator in the $|K_{1,2}\rangle$  basis 
is given by the following $2 \times 2$ diagonal matrix with eigenvalues $0$ and $\GS$:
{\small \begin{eqnarray}\label{augdefwidth}
\widehat{\Gamma}_{\rm WWA} =  \begin{pmatrix} \Gamma-Re(\Gamma_{12}) & 0  \\
0  &  \Gamma+Re(\Gamma_{12}) 
\end{pmatrix} = \gamma  \begin{pmatrix} 0 & 0 \\ 0 & 1 \end{pmatrix},
\end{eqnarray}}
In this case the Lindblad operator $(\ref{lind})$,
related to $\widehat{\Gamma}$ via (\ref{defwidth1}), is given by the following row matrix:
\begin{equation}\label{rowb}
\mathcal{B}~=~\sqrt{\gamma} ( 0~,~1 )~. 
\end{equation}
In the $|K_{1,2}\rangle$ basis (\ref{k1k2}), 
the mass $\widehat{\mathcal{M}}$ matrix, 
which will play the r\^ole of the Hermitian Hamiltonian, is written as~\cite{lopez}:
{\small \begin{equation}
\label{eq:M12}
\widehat{\mathcal{M}}=\left(
\begin{array}{cc}
 M-Re(M_{12}) & -i Im(M_{12}) \\
 i Im(M_{12}) & M + Re(M_{12})
\end{array}
\right)~,
\end{equation}}ignoring again possible CPT-Violating effects. The CP violation parameter
$\epsilon$ is introduced as:
\begin{equation}\label{eq:epsil}
\epsilon =  |\epsilon | \, e^{-i\phi} = \frac{Im (M_{12})}{\frac{\gamma}{2} + i \Delta m}~, \quad {\rm tan}\, \phi = \frac{2 \Delta m}{\gamma} ~,
\end{equation}
where $\Delta m = 2 | M_{12}|$ is the difference between the mass eigenvalues ($m_{1,2} = M \mp |M_{12}|$) of the Kaon mass eigenstates and $\gamma$
is the difference between the width eigenvalues of $K_1$ and $K_2$. At this point we stress that the above parameters, $\Delta m$, $\gamma$ and $\epsilon$, differ from the ones 
of the conventional WWA by terms of order $|\epsilon |^2$. 

The mass eigenstates are found to differ from the $K_{1,2}$ states by terms of order of the CP violation parameter $\epsilon$: 
\begin{eqnarray}\label{eq:eigenstates} 
|\mathcal{M}_1\rangle&= \mathcal{N}_+ \, \Big(i,~\dfrac{Im(M_{12})}{Re(M_{12})+|M_{12}|}  \Big) &=  i |K_1\rangle + \frac{|\epsilon|}{\sin\phi}|K_2\rangle+ O(|\epsilon|^2)  \\
|\mathcal{M}_2\rangle &= \mathcal{N}_- \, \Big(i\dfrac{Re(M_{12}) - |M_{12}|}{Im(M_{12})},~1  \Big) &=   -i\frac{|\epsilon|}{\sin\phi}|K_1\rangle+ |K_2\rangle + O(|\epsilon|^2)~,\nonumber
\end{eqnarray}
where $\mathcal{N}_+ = \frac{Re(M_{12})+|M_{12}|}{\left(2(|M_{12}|^2 + Re(M_{12})|M_{12}|) \right)^{1/2}}$
and $\mathcal{N}_- = \frac{|Im(M_{12})|}{\left(2(|M_{12}|^2 - Re(M_{12})|M_{12}|) \right)^{1/2}}$~are normalization factors.

As this transformation between life-time and  mass eigenstates connects two orthogonal bases,
it is unitary, as shown clearly in Eq.~(\ref{eq:eigenstates}).
The existence of this unitary transformation is a consequence of the incompatibility of $\widehat{M}$ and $\widehat{\Gamma}$ matrices, 
which is described by the invariant determinant of the commutator:
\begin{equation}\label{eq:detmg}
{\rm Det} \Big( \big[ {\widehat {\mathcal M}}\, , \, \widehat{\Gamma} \, \big] \Big) = \Big(2 Re( \epsilon ) \, \Big( (\Delta m)^2  + \frac{\GS^2}{4} \Big)\Big)^2~.
\end{equation}
Notice that the mass-width commutator in (\ref{eq:detmg}) vanishes when $Re(\epsilon)=0$, i.e., when the width difference $\gamma = 0$ or in the absence of  CP violation in the mixing (which implies $Im M_{12} =0$).

In spite of this incompatibility, the WWA treatment of the problem follows the path of diagonalizing 
the entire hamiltonian $\widehat{\mathcal{H}}$ which is not a normal operator.
In this latter case, the eigenvalues are complex with real and imaginary parts differing
from the mass and width eigenvalues by terms of order $|\epsilon|^2$. The corresponding eigenvectors
are the well-known $K_S, K_L$ states, which are not orthogonal as a consequence of Eq.~(\ref{eq:detmg}):
\begin{equation}\label{klks}
\langle K_L | K_S \rangle = 2 Re(\epsilon) ~.
\end{equation}


To solve the evolution equations (\ref{eq:lind}) 
we shall 
use a perturbation method~\cite{lopez}, by which we expand 
the density matrix elements at any time $t$  in powers of the absolute value of the small CP-violation parameter $|\epsilon|$ (\ref{eq:epsil}):
\begin{equation}\label{pert}
\rho_{IJ} (t) = \sum_{n=0}^\infty \, \rho^{(n)}_{IJ}(t) \, |\epsilon|^n~, \quad n \in \mathbb{N}~,
\end{equation}
where the indices $I, J$ span the full Hilbert space of states $\{ i, f \}$, including the decay product (final) state.

In our analysis we shall restrict ourselves to second order in $|\epsilon|$, 
which matches the currently expected experimental sensitivity. Higher orders in $\epsilon$ are complicated and cumbersome to give, leaving aside the fact of not being physically illuminating. On the other hand, keeping terms of  $O(|\epsilon|^2) $ is important, because this is the lowest non-trivial order at which differences between our parameters and those of WWA show up. From  equations (\ref{eq:lind}), the first one  solves the evolution equation 
for the initial states $\rho_{ii'}(t)$, $i, i' = \{1,2\}$, in the $|K_1\rangle$ $|K_2\rangle$ basis, to order $|\epsilon|^2$, and then  
obtain ${\dot \rho_{ff}}(t)$, associated with the $f=\big(\pi, \pi \big)$ decay channel, through 
\begin{equation}
\label{eq:ratff}
{\dot \rho_{ff}}(t) = \gamma \, \rho_{22}(t)~.
\end{equation} 
Notice that with our simplified choice (\ref{rowb}) of  the Lindblad operator $\mathcal{B}$ for a single decay channel, the rate
to this final state is proportional to the probability of having the state $|K_2\rangle$ at time $t$.

The result, expressed in terms of the initial conditions for $\rho_{ii'}(0)$, reads:  
{\small \begin{eqnarray} 
&& \rho_{22}(t) =  \rho_{22}(0)e^{-\GS t}  \nonumber   \\  && -2 |\epsilon| |\rho_{12}(0)| \Big[e^{-\GS t}\cos(\phi+\phi_{12}) 
 -e^{-\frac{\GS}{2} t}\cos(\Delta m t - \phi -\phi_{12}) \Big]   \nonumber \\ && + |\epsilon|^2 \Big[\rho_{11}(0) +  e^{-\GS t }\Big(\rho_{11}(0)+\rho_{22}(0)\big(2\cos(2\phi)+ \GS t\big)\Big)  \nonumber \\ && - 2 e^{-\frac{\GS}{2} t }\, \Big( \rho_{11}(0)\cos(\Delta m t)+\rho_{22}(0)\cos(\Delta m t-2\phi)\Big) \Big] \end{eqnarray} }with $\phi_{12} = {\rm Arg}\rho_{12}(0)$.

We can use this result in order to calculate various observables of the Kaon system, 
in the above approximation of non-decaying $K_1$ state, and compare them with the corresponding ones within the WWA formalism. 
We can build useful observables associated with the decay to $\pi\pi$ or semileptonic decays $\pi l \nu$.
For our purposes here 
we shall concentrate  on two specific observables, namely the decay rates 
$R(K^0 \rightarrow \pi \pi)$ and $R({\overline K}^0  \rightarrow \pi \pi)$.
To be more specific, we will construct separately the sum of rates, sensitive to even powers of $|\epsilon|$, and their difference (or CP-violating asymmetry), sensitive to odd powers of $|\epsilon|$. 

To this end we need the initially  pure \Kz and \Kzb states, prepared experimentally, in the $|K_{1,2}\rangle$ basis. These are described at $t=0$ in the total Hilbert space $H_{tot}$  by the density matrices:
{\small \begin{equation}
\rho_{\Kz} =\dfrac{1}{2}\left(
\begin{array}{ccc}
 \phantom{-}1 & \phantom{-}1 & \phantom{-}0\\
 \phantom{-}1 & \phantom{-}1 & \phantom{-}0\\ 
 \phantom{-}0 & \phantom{-}0 & \phantom{-}0 \\
\end{array}
\right)~, \quad 
\rho_{\Kzb} =\dfrac{1}{2}\left(
\begin{array}{ccc}
 \phantom{-}1 & -1 & \phantom{-}0\\
 -1 & \phantom{-}1 & \phantom{-}0\\ 
 \phantom{-}0 & \phantom{-}0 & \phantom{-}0 \\
\end{array}
\right)~. 
\end{equation}}
To order $|\epsilon|^2$ we obtain:
{\small \begin{eqnarray}\label{eq:decayrates2}
&& \Delta R \equiv R(\Kz\to\pi\pi) - R(\Kzb\to\pi\pi) = -2|\epsilon|\, \GS e^{-t\GS} \big( \cos\phi-  e^{\frac{\GS}{2} t}\cos(\Delta m t-\phi)\big) \nonumber \\
&& \nonumber \\ 
&& \Sigma R \equiv R(\Kzb\to\pi\pi) + R(\Kz\to\pi\pi) = \nonumber \\
&&= \GS \, e^{-t\GS}\Big[ 1 +  |\epsilon|^2\big(1 + e^{t\GS}+t\GS +  2\cos(2\phi)- 4e^{\frac{\GS}{2} t}\cos(\Delta m t-\phi)\cos\phi
\big) \Big] \nonumber \\
&&= \GS \,\Big[ e^{-t\GS}\Big( 1 +  |\epsilon|^2\big(-1 +t\GS +  4\cos^2\phi\big)\Big)+ |\epsilon|^2- 4|\epsilon|^2e^{-\frac{\GS}{2} t}\cos(\Delta m t-\phi)\cos\phi
 \Big] \nonumber \\
&&= \GS (1-|\epsilon|^2) \Big( e^{-t\GS(1-|\epsilon|^2)}\big( 1 +  |\epsilon|^2 4\cos^2\phi\big)\Big)+ \gamma|\epsilon|^2- 4\gamma|\epsilon|^2e^{-\frac{\GS}{2} t}\cos(\Delta m t-\phi)\cos\phi \nonumber \\
\end{eqnarray}}
In our approach,  where $\Delta m$ and $\GS$ are viewed as physical parameters, the time evolution for 
this transition comes from the mismatch  between the basis consisting of states $|K_{1,2}\rangle$ with definite life-times
and the basis of stationary states $|\mathcal{M}_{1,2}\rangle$  with definite mass, as implied by the unitary 
matrix~Eq.~(\ref{eq:eigenstates}).
This mismatch is a consequence of their incompatibility, condensed in the non-vanishing commutator~Eq.~(\ref{eq:detmg}).
For a comparison of the result with WWA, $\Sigma R$ is rewritten as in the last line of Eq.~(\ref{eq:decayrates2}).
To order $|\epsilon|^2$, one notes the appearance of ``\emph{effective widths}'' with values 
$\Gamma_S=\GS (1-|\epsilon|^2)$ and $\Gamma_L=\GS|\epsilon|^2$.
These are precisely the imaginary parts of the complex eigenvalues of the total hamiltonian in WWA. 
The difference between the widths and the ``\emph{effective widths}'' has therefore 
to be taken into account in order to reproduce the same result in both approaches. 

Such corrections between the masses and the real part of the complex eigenvalues
of the total hamiltonian are not seen to order $|\epsilon|^2$, because, as becomes evident from the expressions for $\Delta R$ and $\Sigma R$ in Eq.~(\ref{eq:decayrates2}),
the appearance of $\Delta m$ occurs at least at order $|\epsilon|$. Moreover, the expression of the CP 
violating parameter $\epsilon$ in~Eq.(\ref{eq:epsil}) differs from the one given by WWA, through
the different values for $\Delta m$ and $\Delta \Gamma$ between the two approaches. These differences induce
a relative $|\epsilon|^2$ correction to the complex parameter $\epsilon$ itself.

For those readers concerned by a possible competition of the effects of order $|\epsilon|^2$ with those coming from direct CP violation of order  $\epsilon'$ we point out that what we call ``$\big(\pi \pi\big)$'' in this work  denotes the combination of rates $\frac{1}{3}\big[ 2 (\pi^+ \, \pi^- ) + (\pi^0 \, \pi^0) \big]$, in which the contributions linear in $\epsilon '$ cancel out. 

To conclude, we have presented a description of the decaying neutral Kaon system as an open Lindblad 
system involving evolution of pure to mixed states. It satisfies all the physical requirements of a 
probabilistic quantum mechanical interpretation and guarantees Unitarity, provided that the width operator
 in the dynamics of the initial states is a composite operator expressed in terms of the transition operator $\mathcal{B}$
 between initial and final Hilbert spaces. 
Even if the states $K_{L,S}$ are not physical states filtered by observables,
because there is no measurement associated with the $\Gamma_S$ and $\Gamma_L$ 
effective values, we  have demonstrated 
the equivalence of the observables rates between our approach 
and the Weisskopf-Wigner Approach  to order $|\epsilon|^2$.
It remains to be seen whether a more complete treatment of the Lindblad operator $\mathcal{B}$ 
going beyond the row matrix for a single decay channel, which leads to the factorized dynamics
for the rate Eq.(~\ref{eq:ratff}), would still give this equivalence between the 
two approaches.  
We would like to stress once 
more that the important feature of the Lindblad approach to the neutral-Kaon system is the 
avoidance of using the non-orthogonal $|K_L\rangle $, $|K_S\rangle$ states as a ``\emph{stationary}'' basis. 
We have thus proven that a consistent probabilistic description for the neutral Kaon system exists.

\section*{Acknowledgments}  

One of the authors (PVP) wishes to thank the Physics Department of King's College London for hospitality during the initial stages of this work. This work is supported by the Grants Spanish MINECO  FPA 2011-23596 (JB and PVP),  the Generalitat Valenciana PROMETEO - 2008/004 (JB) and  by the London Centre for Terauniverse Studies (LCTS), using funding from the European Research
Council via the Advanced Investigator Grant 267352 (NEM).

\end{document}